\documentclass[preprint,amssymb, amsmath,%
 aip,cha]{revtex4-1} 

\usepackage{algorithm}
\usepackage{tipa}
 \usepackage{algpseudocode}
 \usepackage{subfigure}
 \usepackage{epstopdf}
\usepackage[usenames]{color}
\usepackage{psfrag}
\usepackage{color}
\usepackage{graphicx}
\usepackage{docs}%
\usepackage{bm}%
\usepackage[colorlinks=true,linkcolor=blue]{hyperref}%
\expandafter\ifx\csname package@font\endcsname\relax\else
 \expandafter\expandafter
 \expandafter\usepackage
 \expandafter\expandafter
 \expandafter{\csname package@font\endcsname}%
\fi
\begin{document}

\title{Constant Communities in Complex Networks}%

\author{Tanmoy Chakraborty}%
\email{its\_tanmoy@cse.iitkgp.ernet.in}
\affiliation{Dept. of Computer Science \& Engg., Indian Institute of Technology, Kharagpur, India -- 721302}%

\author{Sriram Srinivasan}%
\email{ssrinivasan@unomaha.edu}
\affiliation{Dept. of Computer Science, University of Nebraska, Omaha, Nebraska 68106}%

\author{Niloy Ganguly}%
\email{niloy@cse.iitkgp.ernet.in}
\affiliation{Dept. of Computer Science \& Engg., Indian Institute of Technology, Kharagpur, India -- 721302}%

\author{Sanjukta Bhowmick}%
\email{sbhowmick@unomaha.edu}
\affiliation{Dept. of Computer Science, University of Nebraska, Omaha, Nebraska 68106}%

\author{Animesh Mukherjee}%
\email{animeshm@cse.iitkgp.ernet.in}
\affiliation{Dept. of Computer Science \& Engg., Indian Institute of Technology, Kharagpur, India -- 721302}%


\maketitle

Identifying community structure is a fundamental problem in network analysis. Most community detection algorithms are based on optimizing a
combinatorial parameter, for example modularity. This optimization is generally NP-hard, thus merely changing the vertex order can alter
their assignments to the community. However, there has been very less study on how vertex ordering influences the results of the community
detection algorithms. Here we identify and study the properties of {\em invariant} groups of vertices ({\em constant communities}) whose
assignment to communities are, quite remarkably, not affected by vertex ordering. The percentage of constant communities can vary across
different applications and based on empirical results we propose metrics to evaluate these communities. Using constant communities as a
pre-processing step, one can significantly reduce the variation of the results. Finally, we present a case study on phoneme network and
illustrate that constant communities, quite strikingly, form the core functional units of the larger communities.\\
\begin{center}
\line(1,0){475}
\end{center}

A fundamental problem in understanding the behavior of complex networks is the ability to correctly detect
communities. Communities are groups of entities (represented as vertices) that are more connected to each other
as opposed to other entities in the system. Mathematically, this question can be translated to a combinatorial
optimization problem with the goal of optimizing a given metric of interrelation, such as modularity or
conductance. The goodness of community detection algorithms (see~\cite{lf2009,pom2009} for a review) is often
objectively measured according to how well they achieve the optimization.

However, these algorithms can be applied to any network, regardless of whether it possesses a community
structure or not. Furthermore when the optimization problem is NP-hard, as in the case of
modularity~\cite{ng2002}, the order in which vertices are processed as well as the heuristics can change
the results. These inherent fluctuations of the results associated with modularity have long been a source of concern
among researchers. Indeed the goodness of modualrity as an indicator of community structure has also been
questioned, and there exist examples~\cite{gmc2010} which demonstrate that high modularity does not always
indicate the correct community structure. Consequently, orthogonal metrics, such as conductance~\cite{lldm2008}
(which is also NP-complete~\cite{brandes2008}) have been proposed.

Research in addressing the fluctuations in the results due to modularity maximization heuristics include identifying
stability among communities from the consensus networks built from the successive iterations of a non-deterministic
community detection algorithm (such as by Seifi et al.~\cite{sjri2012}). Lancichinetti et al.~\cite{lf2012}
proposed consensus clustering by reweighting the edges based on how many times the pair of vertices were
allocated to the same community, for different identification methods. Delvenne et al.~\cite{dyb2010} introduced the notion of the
stability of a partition, a measure of its quality as a community structure based on the clustered auto-covariance of a dynamic Markov
process taking place on the network. Lai et al.~\cite{lln2010} proposed a random
walk based approach to enhance the modularity of a community detection algorithm. Ovelgonne et
al.~\cite{ogy2012} pointed out an ensemble learning strategy for graph clustering. Gfeller et al.~\cite{gcr2012}
investigated the instabilities in the community structure of complex networks. Finally, several pre-processing
techniques~\cite{scb2012, reidy} have been developed to improve the quality of the solution. These methods form
an initial estimate of the community allocation over a small percentage of the vertices and then refine this
estimate over successive steps.

All these methods focus on compiling the differences in the results to arrive at an acceptable solution, and despite these advances a
crucial question about the variance of results remains unanswered -- what do the {\it invariance} of the results tell us about the network
structure? In this paper, we focus on the invariance in community detection as obtained by modularity maximization. Our results, on a set of
scale-free networks, show that while the vertex orderings produce very different set of communities, some groups of vertices are always
allocated to the same community for all different orderings. We define the group of vertices that remain invariant as {\it constant
communities} and the vertices that are part of the constant communities as {\it constant vertices}. Figure 1 shows a schematic diagram of
constant communities. Note that not all vertices in the network belong to constant communities. This is a key difference of constant
communities with the consensus methods~\cite{lf2012} described earlier. Consensus methods attempt to find the best (most stable or most
similar) community among all available results and thus include all the vertices. Constant communities, on the other hand, focus on finding
subgraphs where the cohesive groups can be unambiguously identified. As discussed earlier, communities obtained by modularity maximization
may
include vertices that can move from one group to another depending on the heuristic or the vertex ordering. The vertex groups obtained using
constant communities are invariant under these algorithmic parameters and, thereby, provide a lower bound on the number of uniquely
identifiable communities in the network. Although trivially each vertex can be considered to be a
constant community by itself, our goal is to identify the largest number of vertices (i.e., at least three or more) that can be included in
an invariant group.

\begin{figure}
\centering
\includegraphics[scale=0.3]{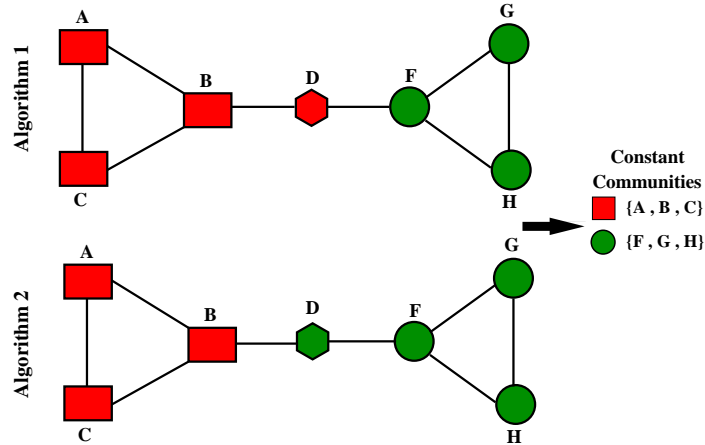}
\caption{ {\bf Schematic illustration of the formation of constant communities.} Two colors (red and green) indicate two communities of
the network formed in each iteration. Combined results of two algorithms produce two constant communities (rectangular and circular
vertices).
Remaining one vertex (hexagonal shaped) is not included since it switches its community between the two algorithms.}
\end{figure}

The presence of such invariant structures can be used to evaluate the accuracy of the communities when other independent methods of
verifications are unavailable.
However in many networks, constant communities constitute only a small percentage of the total number of vertices. To understand how other
non-constant vertices are allocated to communities, we show that by using constant communities we can significantly reduce the variations in
results. Thus, building from the more accurate results reduces the variance over the larger network. In
brief our main
contributions are as follows:

\begin{itemize}
\item demonstrate the possibility of extreme variance in community structure due to vertex perturbations
\item develop metrics to determine whether a network possess invariant groups of constant communities
\item demonstrate how using constant communities as a pre-processing step can reduce the variance in modularity maximization methods.
\end{itemize}

\section*{Results}
\textit{Experimental setup}. In this section, we first demonstrate that even for the same optimization objective (in this case
maximizing modularity) and the same heuristic, the inherent non-determinism of the method can significantly change the results. Based on
our results, we define metrics to estimate the propensity of a network to form communities. Finally, we show how
combining constant communities as a pre-processing step can help improve the modularity of the community detection algorithm for the network
as a whole.

We selected two popular agglomerative modularity maximization techniques -- the method proposed by Clauset et al.~\cite{cnm2004}
(henceforth referred to as the CNM method) and the method proposed by Blondel et al.~\cite{bgll2008} (henceforth referred to as Louvain
method). Both these methods initially start by assigning one vertex per community. Then at each iterative step, two communities whose
combination most increases the value of modularity are joined. This process of joining community pairs is continued until the value of
modularity no longer increases. The Louvain method generally produces a higher value of modularity than CNM, because it allows vertices to
migrate across communities if that leads to a more optimum value.

In order to identify these
communities, for each network in the test suite, we applied the CNM (and Louvain) method over different permutations of the vertices and
then
isolated the common groups that were preserved across the different orderings (see Methods section). These common groups of vertices were
marked as the constant communities for the respective
network.

We identified constant communities using both the CNM and Louvain algorithms. We observed based on the high ($>$ 0.80) Normalized Mutual
Information (NMI)~\cite{nmi} (see the supplementary information for the definition of NMI) values that the overlap
between the constant communities obtained from the two methods is considerable~\cite{nmi_1, nmi_2} (see Table III in the supplementary
information). Therefore, in the interest of space and
clarity we confine our discussion about the properties of constant communities to those obtained from the Louvain method.

{\it Degree preserving order.} Ideally, the total number of different orderings to be tested should be equal to the factorial of the number
of vertices in the network. However, even for the smallest network in our set (Chesapeake with 39 vertices) this value is astronomical. We
therefore restrict our permutations to maintain a {\it degree-preserving }order. The vertices are ordered such that if degree of $v_i$ is
greater than the degree of $v_j$, then $v_i$ is processed prior to $v_j$.

In addition, to reducing the number of vertex permutation, degree-preserving permutation also has another important advantage. Recall
that
the networks in the test suite have few vertices with high degrees and a lot with low degrees. Therefore, arranging the high degree vertices
earlier pushes most of the fluctuations towards the later part of the agglomeration process. This ensures
that the sub-communities formed initially are relatively constant and only later do the divergence in community memberships take place.
Clearly, such orderings based on decreasing degrees are geared towards facilitating low variance in communities. If even {\it this ordering}
does not produce constant structures, it makes a very strong case about the inherent fluctuations that underlie modularity
maximization methods.

\textit{Test suites}. Our experiments were conducted on networks obtained from real-world data as well as on a set of synthetically
generated networks using the LFR model~\cite{lfr2009}. The set of real-world networks is obtained from the instances available at the
10th DIMACS
challenge website~\cite{dimacs}. The networks, which are undirected and unweighted, include -- Jazz (network of jazz musicians;
$|V|=198, |E|=2742$)~\cite{jazz}, Polbooks (network of books on USA
politics; $|V|=105, |E|=441$)~\cite{polbooks}, Chesapeake (Chesapeake bay mesohaline network; $|V|=39, |E|=340$)~\cite{chesapeake},
Dolphin (Dolphin social network; $|V|=62, |E|=159$)~\cite{dolphin}, Football (American college football; $|V|=115,
|E|=1226$)~\cite{football}, Celegans (Metabolic network of C. elegans; $|V|=453, |E|=2025$)~\cite{celegans}, Power (topology of the Western
States Power Grid of the USA; $|V|=4941, |E|=6594$)~\cite{power} and Email (e-mail
interchanges between members of the Univeristy Rovira i Virgili; $|V|=1133, |E|=5451$)~\cite{email} (note that $|V|$ refers to the number
of vertices and $|E|$ refers to the number of edges). All these networks
exhibit scale-free degree distribution (see Figure S1 in the supplementary information).

Networks generated using the LFR model are associated with a mixing parameter $\mu$ that represents the ratio of the external connections
of a node to its total degree. We created LFR networks based on the following parameters~\cite{lf2012}: number of nodes = 500, average
degree = 20,
maximum
degree = 50, minimum community size = 10, maximum community size = 50, degree exponent for power law = 2, community size exponent = 3. We
varied the value of $\mu$ from 0.05 - 0.90. Low values of $\mu$ correspond to well-separated communities that are easy to detect and
consequently these networks contain larger percentage of constant communities. As $\mu$ increases, communities get more ambiguous and
community detection algorithms provide more varied results leading to fewer vertices being in significantly sized constant communities.

\textit{Sensitivity of community structure to vertex perturbations.} In our first experiment we study how the community structures of the
networks change under vertex perturbations. Since constant communities are the groups
of vertices that remain invariant, we measure the change in community structure based on the number of constant
communities. We define \textit{sensitivity} ($\phi$) as the ratio of the number of constant communities to the
total number of vertices. If $\phi$ is 1 then each vertex by itself is a constant community (the trivial case),
thus there is no consensus at all over the set of communities obtained over different permutations. The higher
the sensitivity metric, the fewer the vertices in each constant community and, therefore, this metric is useful
for identifying networks that do not have a good community structure under modularity maximization.

\begin{figure}
\centering
\includegraphics[scale=0.33]{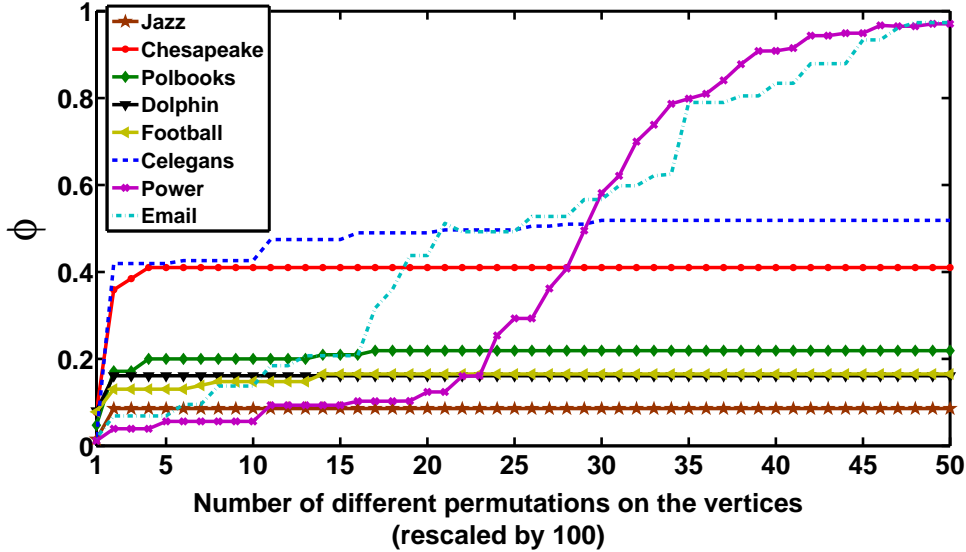}
\caption{ {\bf Sensitivity of each network across 5000 permutations.} X-axis indicates the number of
permutations. The x-axis is rescaled by a constant factor of 100 for better visualization. Y-axis indicates the value of
sensitivity as it changes over the permutations. Power and Email networks have very high sensitivity values
indicating that they possibly do not have a tightly knit community structure.}\label{sensitivity}
\end{figure}

The sensitivity of each network is given in Figure~\ref{sensitivity}. The x-axis indicates the number of
different permutations of the vertices and the y-axis plots the value of the sensitivity. We observe that for
most of the networks the number of constant communities become stable within the first 100 permutations, and the
sensitivity values are low. This indicates that there can potentially exist very strong groups in these networks
that have to be together to achieve high modularity. However, for networks such as Power grid and Email,
the number of constant communities kept increasing until the values of $\phi$ were close to 1. Thus, the
community detection results for these two networks are extremely sensitive to the vertex perturbations. This
implies that the communities (if any) in these two networks are not tightly knit, i.e., very ``amorphous''.

\textit{Percentage of constant communities}. We now investigate, in further detail, the properties of constant communities. We define the
\textit{relative size} ($\xi$) of a constant community as the ratio of the number of vertices in that constant community to the total number
of vertices in
the network and the \textit{strength} ($\Theta$) as the ratio of the edges internal to the constant community to the edges external to the
constant community.

\begin{figure}[!h]
\centering
\includegraphics[scale=0.33]{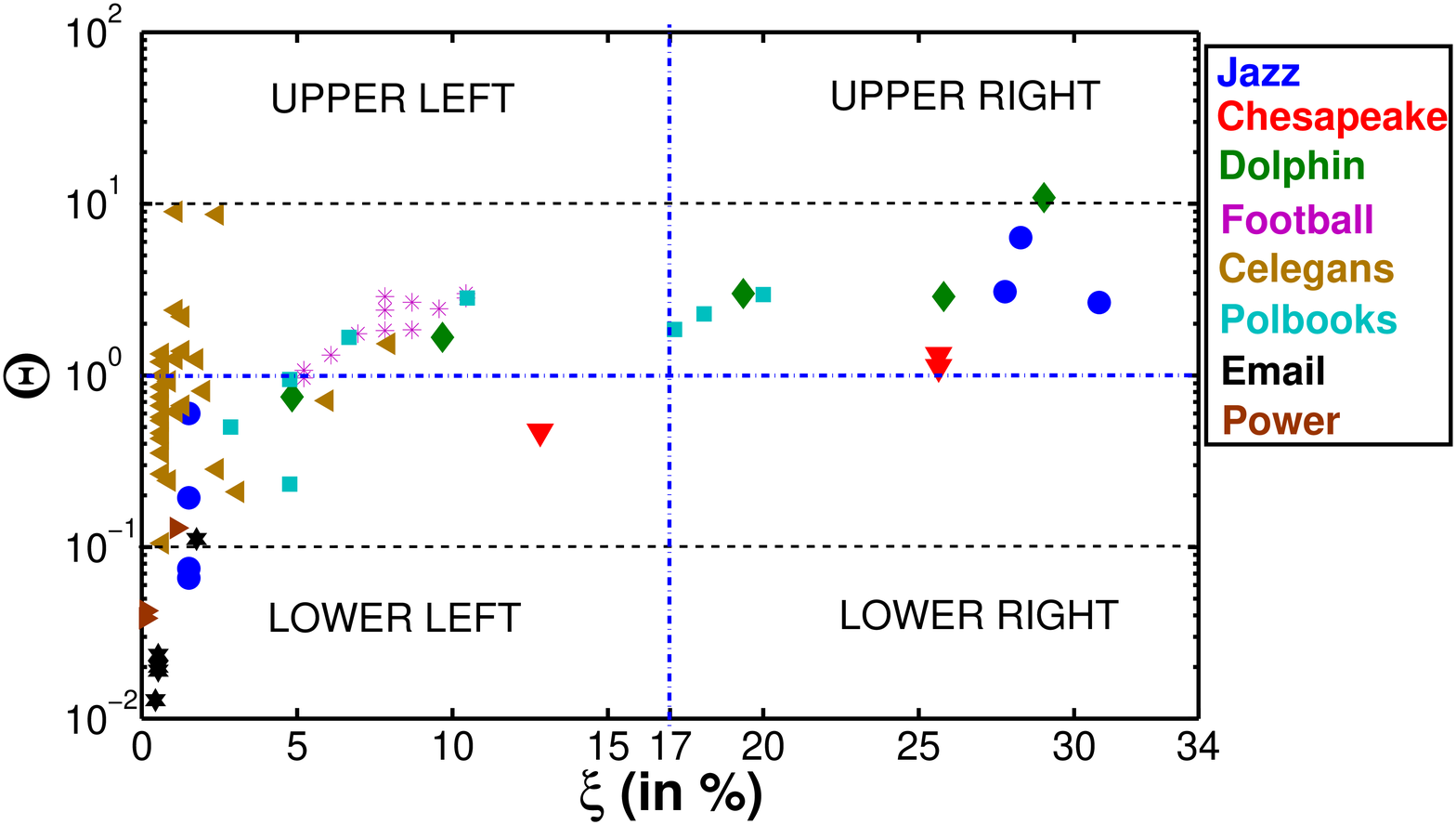}
\caption{ {\bf Comparison between the relative size and strength of the constant communities}. X-axis plots the
relative size in percentage. Y-axis (in logarithmic scale) plots the strength. Jazz, Dolphin, Polbooks and
Chesapeake show strong constant community structure. But Email and Power hardly have any constant communities.
The plot is vertically divided at x = 17 that could help systematically analyze the distribution of the points.
}\label{intr_extr}
\end{figure}

Figure~\ref{intr_extr} plots the relative size (in percentage) of the constant communities with respect to their
strength. If the strength of a constant community is above 1 (above 0 in log scale) then the number of internal
edges in the community is larger than the number of external edges. The higher the value, the more tightly
connected is the community. We see that the value of relative size ranges from 0-34, with a larger cluster of
values around 0-5. This shows that most of the constant communities contain very few vertices with respect to
the network. If the relative size of the constant communities is low then the remaining vertices have more
freedom in migrating across communities, making the community structure weaker. We observe that, despite there
being more constant communities of low relative size, there are some networks that have multiple constant
communities with relative size over 15\% of the total number of nodes indicating that they have a much stronger
community structure. These include Jazz, followed by Dolphin and then Polbooks and Chesapeake.

Relative size and strength together provide an estimate of which networks have good community structure. If we divide the x-axis at
roughly the
mid-point of the range and the y-axis at 1, then we obtain four quadrants each representing different types of community structures.
The first quadrant (upper right)
contains communities that have high relative size as well as high strength. Networks containing a large number of such constant communities
are less likely to be affected by perturbations. Diagonally opposite is the third quadrant (lower left), which contains communities of low
relative size and low strength. As discussed earlier, networks having communities predominantly from this quadrant will produce
significantly
different results under perturbations and are likely to not have a strong community structure under modularity maximization. The second
quadrant (upper left) contains the
groups of vertices that are strongly connected but have small relative size. This indicates that there are some pockets of the network with
strong community
structure. The fourth quadrant (lower
right) represents communities with high relative size but low strength. In this set of experiments it is empty, and we believe that
this area will be sparsely populated, if at all. This is because networks having such communities will have a very special structure:
strongly connected groups of very few vertices with many spokes radiating out to account for the high number of external communities.

\textit{Pull from external connections}.
We note in Figure~\ref{intr_extr} that there are several constant communities whose strength is below one,
i.e., they have more external than internal connections. This is counterintuitive to the idea that a strong community should have more
internal connections. Indeed, modularity maximization methods always tend to create communities whose strengths are greater than
one. However, the structure of some of the constant communities belies this convention.

\begin{figure}[!h]
\centering

\subfigure{
\includegraphics[scale=0.373]{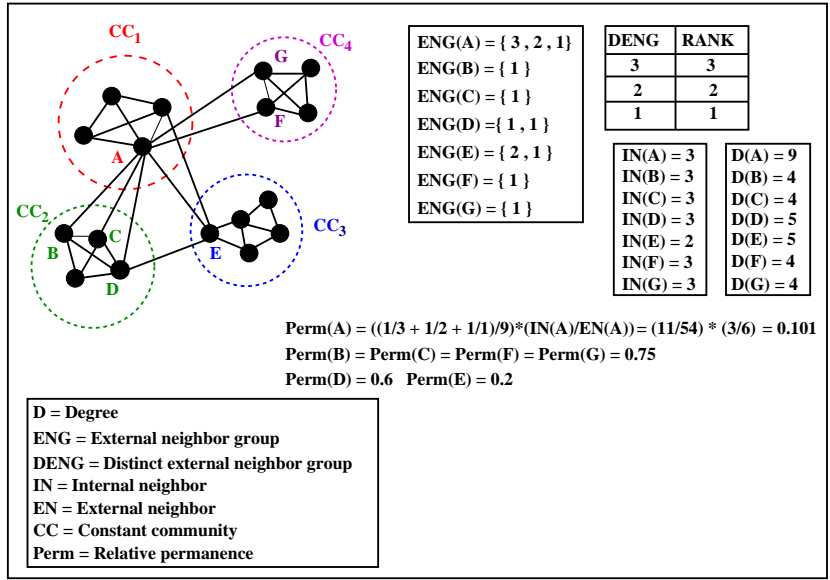}}
\subfigure{
\psfrag{omega}{{\bf $\Omega$}}
\psfrag{P (omega)}{{\bf P$(\Omega)$}}
\includegraphics[scale=0.3] {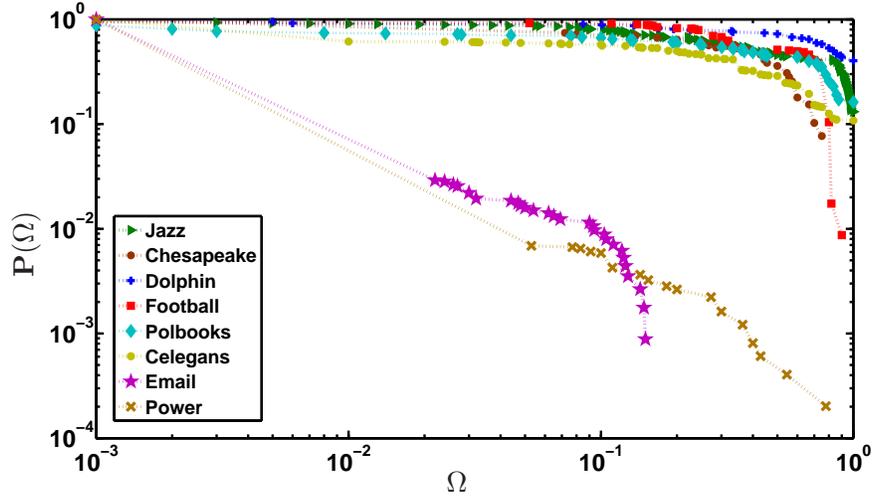}}
\caption{ {\bf Top: Schematic diagram illustrating the computation of the relative permanence of the
vertices}.
{\bf Bottom: Distribution of relative permanence values}. X-axis plots
the value of $\Omega$ and y-axis plots the cumulative fraction of vertices ($P(\Omega)$) exhibiting that $\Omega$. Both axes are in
logarithmic
scale.}\label{pull}
\end{figure}

We observe that in these cases, the external connections are distributed across different communities. Furthermore, the number of
connections to
any one external community is always lower than the internal connections. Based on this observation, we hypothesize that a group of
vertices are likely to be placed together so long as the internal connection is greater than the connections to any one single external
community. Then the vertices within the community do not experience a significant ``pull'' from any of the external communities that will
cause them to migrate, and, therefore, their propensity to remain within their own communities is high. We quantify this measurement as
follows:

Let $v$ be a vertex in a constant community; further, let $D(v)$ denote the degree of $v$, and $EN(v)$ and
$IN(v)$ denote the number of external and internal neighbors of $v$ respectively (i.e., $D(v)=IN(v)+EN(v)$). We
also assume that the $EN(v)$ external neighbors are divided into $k$ external groups, and $ENG(v)$ denote a set
of $k$ elements where the \textit{i}th element in the set represents the number of neighbors of $v$ belonging to
the \textit{i}th external group. For instance, considering the vertex $A$ in $CC_1$ in Figure~\ref{pull} (Top),
$D(A)=9, IN(A)=3, EN(A)=6$ and $ENG(A)=\{3,2,1\}$ (i.e., three external neighbors in $CC_2$, one external
neighbor in $CC_3$, and two external neighbors in $CC_4$). Similarly, we calculate $ENG(v)$ for each vertex in
the network and form a list $DENG(G)$ by taking union over all $ENG(v)$, that is, only  unique entries across
$ENG(v)$ get listed in $DENG(G)$ (see Figure~\ref{pull}-top). The list is then ranked in ascending order.
The intuition behind this ranking is to identify the diverse range of the sizes of the external groups. The inverse
of the rank would therefore signify the intensity of the pull of the particular external community. For a
particular vertex, if the inverse rank of each of the external group is equal to one, it would point to the fact
that all its external neighbors are diversely distributed (i.e., well-sparsed), and therefore the pull
experienced should be minimum; in contrast, if the value is much lower than one, it would imply that the  vertex
experiences a strong pull from its external neighbors. We define the {\it strength} of a vertex {\it v},
$\theta(v)$, as the ratio of the internal neighbor ($IN(v)$) to the external neighbor ($EN(v)$) of vertex $v$
similar to  the strength ($\Theta$) of a constant community defined earlier. Mathematically, the suitably
normalized value of $relative\ permanence$, $\Omega(v)$, of a vertex $v$ in a constant community can be expressed as:

\begin{equation}
\Omega(v)= \theta(v) \times \frac{\sum_{i=1}^k {\frac{1}{Rank_i(ENG(v))}}}{D(v)}
\end{equation}

where $Rank_i(ENG(v))$ denote the rank (retrieved from the {\it DENG(v)} list) of the \textit{i}th element in $ENG(v)$. This metric
indicates the
propensity of a vertex to remain
in the same community regardless of
any algorithmic parameters.

Figure~\ref{pull} (Top) presents a schematic diagram for computing relative permanence of vertices within the communities. Figure 4 (Bottom)
plots the
cumulative
distribution of the relative permanence over the vertices in all networks. The x-axis indicates the value of the relative permanence and the
y-axis, the cumulative fraction of vertices having the corresponding relative permanence value.
The nature of the cumulative permanence distribution of the vertices is roughly same for all networks except Email and Power. The
distinguishing nature of the curves for Email and Power graphs compared to the other graphs indicates that very few number of vertices in
these two networks have higher relative permanence values and therefore experience more ``pull'' from the external communities.
 Another observation is that a high fraction of
vertices in Jazz, Polbooks, Dolphin and Celegans have
relative permanence close to one. These vertices are more ``stable'' compared to the other vertices in the respective networks.

\textit{Constant communities for improving the modularity}.
 We note that in many networks (such as Football and Celegans) constant communities form only a small percentage of the vertices. Thus,
finding only the constant communities may not provide adequate information about the relationship amongst the
rest of the vertices. We therefore leverage on the invariant results in the first and second quadrants of
Figure~\ref{intr_extr} as building blocks to identify larger communities.

\begin{table}
\caption{Modularity before and after pre-processing for real networks (left) and for different values of mixing parameter ($\mu$) over LFR
graphs (right)}\label{modularity}
\parbox{.49\linewidth}{
\centering
\scalebox{0.7}{
\begin{tabular}{|c|c|c|c|c|}
 \hline
 & \multicolumn{4}{|c|}{Louvain}\\
\cline{2-5}
Networks & \multicolumn{2}{|c|}{Before pre-processing} & \multicolumn{2}{|c|}{After pre-processing} \\
\cline{2-5}
&  Mean ($m_q$) & Var ($\sigma_q$) & Mean ($m_q$) & Var ($\sigma_q$) \\
\hline
Jazz &  0.448 & 3.13e-6 & 0.452 & 0 \\
\hline
Chesapeake &  0.301 & 1.17e-5 & 0.303 & 3.36e-33 \\
\hline
Polbooks &  0.539 & 1.74e-5 & 0.557 & 1.24e-32 \\
\hline
Dolphin & 0.543 & 1.76e-5 & 0.550 & 0 \\
\hline
Football &  0.610 & 2.01e-5 & 0.623 & 0 \\
\hline
Celegans & 0.438 & 2.89e-5 & 0.442 & 1.33e-26 \\
\hline
Email & 0.542 & 6.89e-5 & 0.568 & 0.95e-12 \\
\hline
Power & 0.936 & 1.09e-5 & 0.937 & 2.25e-10 \\
\hline
\end{tabular}}
}
\hfill
\parbox{.49\linewidth}{
\centering
\scalebox{0.775}{
\begin{tabular}{|c|c|c|c|c|c|}
  \hline
 & &\multicolumn{4}{|c|}{Louvain} \\
\cline{3-6}
$\mu$& Planted&\multicolumn{2}{|c|}{Before } & \multicolumn{2}{|c|}{After}\\
 & Modularity& \multicolumn{2}{|c|}{pre-processing} & \multicolumn{2}{|c|}{pre-processing} \\
\cline{3-6}
&  &Mean($m_q$) & Var($\sigma_q$) & Mean($m_q$) & Var($\sigma_q$) \\
\hline
0.05 & 0.878 & 0.834 & 1.98e-24 & 0.877 & 0 \\
\hline
0.10 & 0.817 & 0.802& 2.28e-28& 0.817& 0\\
\hline
0.20 & 0.716 & 0.690& 5.74e-7& 0.686& 0\\
\hline
0.50 & 0.440 & 0.385& 2.05e-6 & 0.389& 1.58e-28\\
\hline
0.70 & 0.223 & 0.298& 9.70e-10& 0.219& 1.04e-28\\
\hline
0.90 & 0.029 & 0.225& 4.25e-10& 0.205& 5.64e-28\\
\hline
\end{tabular}}
}
\end{table}

We first combine the constant communities into super-vertices. This process creates a smaller network as well as ensures
 that the vertices in the constant communities always stay together. Then we execute a modularity maximization algorithm
over the entire network (see Methods section). We compute the variance in results by executing
the
underlying modularity maximization algorithm over 5000 permutations using the degree-preserving order. As shown in Table~\ref{modularity}
(left), combining constant communities as a pre-processing step both increases the mean modularity value as well as reduces the variability
across permutations for real-world networks.

We also observe that the variance becomes 0 or very low for the networks which have significant number of constant
communities in the first and second quadrants of Figure~\ref{intr_extr}. The results obtained from the other
networks with high sensitivity, such as Email and Power, still indicate some variance although the value is less pronounced.

 These observations on real-world networks lead us to believe that pre-processing using constant communities is more effective if a network
has strong community structure. To test this hypothesis, we created
LFR graphs with mixing parameters from 0.05 to 0.90. Low mixing parameters indicate strong community structure. As shown in
Table~\ref{modularity} (right), pre-processing using constant communities helps increase the modularity value and reduces variability of
the results.

\begin{figure}[!h]
\centering
\psfrag{mu}{{\large{ {\bf $\mu$}}}}
\includegraphics[scale=0.3]{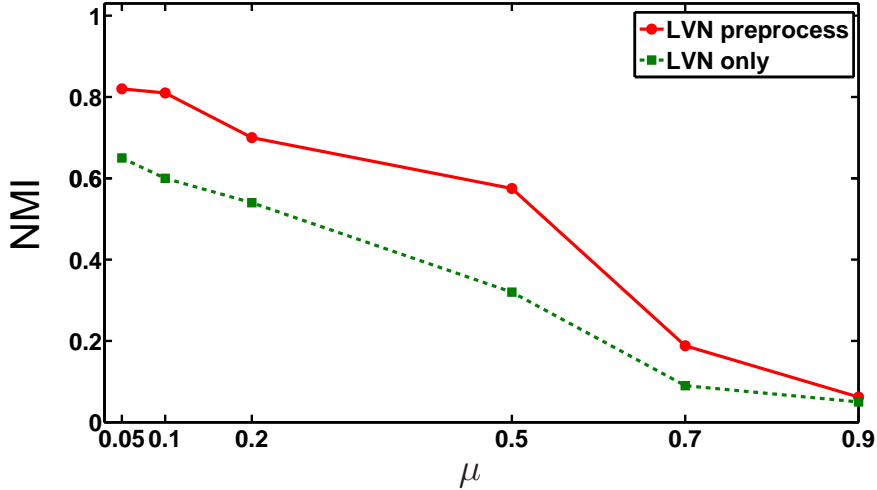}
\caption{ {\bf Variation of NMI for different values of mixing parameters}. The broken line corresponds to the experiment without the
pre-processing step
and the solid line to the experiment after using the pre-processing step.}\label{nmi}
\end{figure}

Another advantage of LFR networks is that we know the ``ground truth'' i.e., the correct distribution of
communities (exact number of vertices in each community and the number of in-community connections between them). We used NMI to compare the
communities obtained, with and without using the pre-processing step. As
shown in Figure~\ref{nmi}, when the community structure is strong (low mixing parameter), using constant
communities pushes the result towards the ground truth. In contrast, when the
community structure is not well-defined (high mixing parameter), use of constant communities does not mimic the
community distribution of the ground truth, because there can be many variations of community distribution in
such networks that lead to high modularity. These results once again highlight the significance of constant
communities.

\textit{Relative ranking of constant communities.} A constant community is strong if it is large (high $\xi$) or is well-connected
(high $\Omega(v)$). We experimented to see  which one of these two properties is more important in determining high modularity. To do so,
we ordered the constant communities according to (a) decreasing order of $\xi$ and (b) decreasing order of $\Omega$. We combined the
constant communities into super-vertices one by one following the order obtained from (a) and (b) separately. After each combination, we
computed the modularity and compared the value with the average modularity (over 5000 permutations) obtained by using the Louvain
method without any pre-processing. 

\begin{figure}[h]
\centering
\includegraphics[scale=0.33] {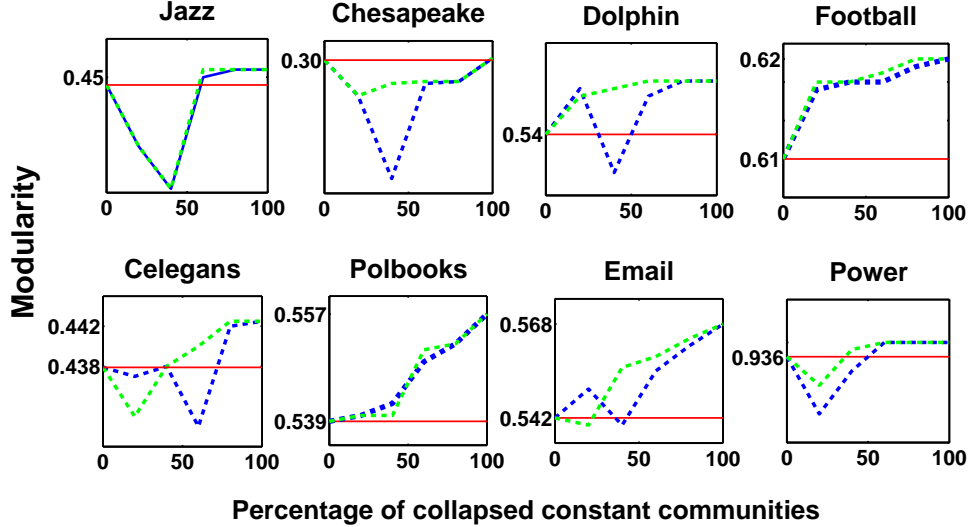}
\caption{ {\bf Modularity after partially collapsing the constant communities}.  The broken blue lines are in decreasing order
of size and the broken green lines are in decreasing order of relative permanence. The red lines depict the mean modularities
without using constant communities.}\label{big}
\end{figure}

Figure~\ref{big} compares the modularity obtained by collapsing constant communities according to the order obtained from (a) (dotted blue
line) and (b) (dotted green lines). For almost all the networks, there is a transition where the modularity values cross over the mean
modularity (solid red line). Once this transition takes place, the modularity
values generally remain above (or at least equal to) the mean modularity.  This critical point indicates the smallest fraction of constant
communities required to outperform the original algorithms. We observe further that
the green lines (ordered according to $\Omega(v)$) generally reach the critical point earlier than the blue lines (ordered according to
 $\xi$), indicating that $\Omega(v)$ is a better indicator of constant communities.

\textit{Case study.} The significance of constant community in a network can be further understood if we consider networks where
nodes have specific functionalities associated with them. We hypothesize
that
in such a network a constant community would represent indispensable functional blocks that reflect the defining characteristics of the
network. In order to corroborate this hypothesis we conduct a case study on a specific type of linguistic network constructed from the
speech sound inventories of the world's language~\cite{pho}. The sound inventory of a language comprises a set of consonants and vowels also
sometimes together known as phonemes. In order to unfurl the co-occurrence principles of consonant inventories, the authors~\cite{pho}
constructed a network (phoneme-phoneme network or PhoNet) where each node is a consonant and an edge between two nodes denotes if the
corresponding consonants have co-occurred in a language. The number of languages in which the two nodes (read consonants) co-occur defines
the
weight of the edge between these nodes. Note that each node here has a functional representation since it can be represented by means of
a set of phonetic features (e.g., bilabial, dental, nasal, plosive etc) that indicate how it is articulated. Since this is a weighted
graph, we suitably define a threshold to construct the unweighted version. We compute constant communities of PhoNet and observe that each
such graph (see Table~\ref{phon}) represents a {\em natural class}, i.e., a set of consonants that have a large overlap of the
features~\cite{pho}. Such groups are frequently found to appear together across languages, and linguists describe this observation through
the principle of {\em feature economy}~\cite{pho}. According to this principle, the speakers of a language tend to be economic in choosing
the features in order to reduce their learning effort. For instance, if they have learnt to use a set of features by virtue of learning a
set of sounds, they would tend to admit those other sounds in their language that are combinatorial variations of the features already
learnt -- if a language has the phonemes /p/ (voiceless, bilabial, plosive), /b/ (voiced, bilabial, plosive) and /t/
(voiceless, dental, plosive) in its inventory then the chances that it will have /d/ (voiced, dental, plosive) is disproportionately
higher compared to any other arbitrary phoneme since by virtue of learning to articulate /p/, /b/ and /t/ the speakers need to learn no new
feature to articulate /d/. Identification of constant communities therefore systematically unfolds the natural classes
and provides a formal definition for the same (otherwise absent in the literature). We plot in
Figure S2 (see supplementary information), the average hamming distance between the feature vectors of phonemes forming a constant
community versus the community size. The average hamming distance is significantly lower in the case when a set of randomly
chosen phonemes are
grouped together and assumed to represent a community with varying sizes as that of the constant communities. Further, we observe that
collapsing the constant communities results either in more dilute groups (still with a certain degree of feature overlap) or reproduces the
same constant communities indicating that no valid dilution is possible for these functional blocks.

\begin{table}
\caption{Few constant communities of PhoNet and the features they have in common}\label{phon}
 \begin{tabular}{|c|c|}
\hline Constant communities & Features in common\\\hline /p\textsuperscript h/, /t\textsuperscript h/, /k\textsuperscript h/& voiceless,
aspirated, plosive\\\hline
/\textsuperscript mb/, /\textsuperscript nd/, /\textsuperscript ng/     & prenasalized, voiced, plosive\\\hline
 /\textsubtilde{p}/, /\textsubtilde{t}/, /\textsubtilde{k}/ & laryngealized, voiceless, plosive\\\hline
/\textipa{t}/, /\textipa{d}/, /\textipa{n}/& dental\\\hline /\textipa{\:l}/, /\textipa{\:n}/, /\textipa{\:t}/, /\textipa{\:d}/&
retroflex\\\hline
 \end{tabular}
\end{table}

\section*{Discussion }
Constant communities are regions of the network whose community structure is invariant under different perturbations and community detection
algorithms. They, thereby, represent the core similar relationships in the network. The existence of multiple results for community
detection is well known; however, this is one of the first studies of the invariant subgraphs that occur in a network.

Although we currently detect constant communities by comparing across different permutations, our results have uncovered
some interesting facets about the community structure of networks, which can lead to improved algorithms for community detection. First, we
observe that constant communities do not always have more internal connections than external connections. Rather, the strength of the
community is determined by the number of different external communities to which it is connected. We have proposed a metric to quantify
the pull that a vertex experiences from the external communities and the relative permanence of the said vertex
indicating its inertia to stay in its own community.

Secondly, in most networks, constant communities cover only a subset of the vertices. Depending on the size of the constant communities it
may not be correct or necessary to assign every vertex to a community, as is the focus of most community detection algorithms. Furthermore,
even if when we insist on assigning a community to each vertex, the constant communities can be leveraged to produce results with higher
modularity and lower variance. Thus, as discussed earlier, constant communities form the smallest indivisible units in the networks and
particularly in the case of agglomerative methods can be used to hierarchically build larger communities.

Thirdly, the high functional cohesion among the vertices of the constant community can render meaning to the community structure of
the
networks. This
conclusion is much more apparent for labeled graphs where the vertices are associated with certain functional
properties. If we stop at
detecting only the constant communities and treat them as the actual community structure of the graph, we observe that sometimes it acts as
a hard bound since no further community detection might be possible. Therefore, we suggest that the prior
detection of these building blocks is always significant in order to further decide to
merge them into more coarse-grained communities pertaining to a diluted functional cohesion.

The fourth and most important observation is that not all networks have significant constant community structure. The two most egregious
examples in our test suites are Power and Email graphs. The absence of constant communities in the networks indicate that either communities
in general do
not exist or they are highly overlapped and therefore do not have a significant constant region. The first case is true for Power grid,
which as a grid is unlikely to have communities. We believe that the second reason probably holds for the Email network. A set of
professional
emails within correspondents in the same university is likely to have more overlaps than clear cut communities.

Finally, we have demonstrated evidence that the modularity measure is not enough to judge the inherent compartmental structure of a network.
For instance, Email and Power networks have reasonably higher modularities compared to the others. Still, no consensus is observed in their
community structures. Rather their sensitivity measures indicate that each node might act as individual constant community in the
further iterations. Therefore, the goodness metric of the community detection algorithm should be redefined in a way that can effectively
capture the modular structure of the network.

We note that the experiments in this paper focused solely on agglomerative modularity maximization methods. We plan to continue our studies
on the effect of vertex perturbations on other types of community detection algorithms such as divisive and spectral methods as well as
different optimization objectives. In particular we are very keen to understand how the randomness of a network could be quantified
in order to develop algorithms that take into account the variation in randomness of connections for determining the quality of the
communities.

\section*{Methods}
\textit{Identifying constant communities.} In order to identify constant communities we permute the order of the vertices, and then apply a
community detection algorithm to each of the permuted networks. The results vary across permutations. We select the groups of vertices that
were always allocated together across all the permutations and mark them as constant communities. Algorithm 1 in supplementary information
formalizes the steps to find out constant communities (see Figure S3 for the schematic diagram of the algorithmic steps in the supplementary
information). The rationale behind this process is that these vertices must have some intrinsic connectivity properties that force them to
stay together under all orderings.

To implement the vertex permutation, we adopt a stochastic degree-preserving scheme that can arrange the vertices based on the descending
order of their degrees.  The ordering of the set of vertices with the same degree is permuted. By applying this method we preserve the
relative ordering of the degrees of the vertices since it is well-known that node-degrees constitute a fundamental network property. We have
also observed that the random
permutations producing high modularity usually preserve a degree-descending order of vertices and the ones that result in low modularity
usually are outcomes of cases where the algorithm would start executing from a low-degree vertex. Thus, our permutations prevent us from the
possibility of getting confined in a local maximum of the modularity.

\textit{Combining constant communities for modularity maximization.} For these tests, we first collapse the constant communities to
individual nodes (see Figure S3 in the supplementary information). This step ensures that the constant vertices
are always grouped together and are guaranteed to remain within the same community. The total
number of edges between the vertices of the two collapsed communities is computed and this sum is assigned as the new edge weight between
them. We then apply a community detection method to the new weighted network to obtain the final modularity.

\bibliographystyle{model1-num-names}

\section*{Acknowledgement}
This work has been supported by the Google India PhD fellowship Grant in Social Computing, College of IS\&T at University of Nebraksa at
Omaha(UNO) and the FIRE and GRACA grants from the Office of Research and Creative Activity at UNO.

\section*{Author contributions}
T.C., S.B., N.G., A.M. designed research; T.C., S.S., S.B., N.G., A.M. performed research; T.C., S.S., S.B., N.G., A.M. contributed new
reagents/analytic tools; T.C., S.S., S.B., N.G., A.M. analyzed data and T.C., S.B., N.G., A.M. wrote the paper.

\section*{Additional information}
Supplementary information accompanies this paper.\\
{\bf Competing financial interests:} The authors declare no competing financial interests.

\newpage
    
\hspace{1.5 in} {\bf Supplementary Information}
\maketitle

\section*{Definitions, formulae and notations}
This section contains the definition of some of the terms used in the main text. Most of the networks considered
here are undirected, unweighted and connected graphs, $G(V,E)$, where $V$ is the set of vertices and $E$ is the
set of edges. An edge $e \in E$ is associated with two vertices ${u,v}$ which are called its endpoints.  A
vertex $u$ is a neighbor of $v$ if they are joined by an edge. $N(v)$ is the set of neighbors of vertex $v$ and the
degree of $v$, $degree(v)$, is equal to $|N(v)|$, the cardinality of the set of its neighbors.

\subsection{Clustering coefficient}
Clustering coefficient measures the propensity of the network to form clusters. The local clustering coefficient of a vertex $v$ is computed
as the ratio of the edges between the neighbors of a vertex to the total possible connections between the neighbors, as follows:\\
\begin{equation}
 C(v)=\frac{2 \times |e_{ij}| }{N(v)\times(N(v)-1)};\ \ i,j \in N(v)
\end{equation}

where $N(v)$ is the set of neighbors of $v$, $e_{ij}$ is the set of edges between the neighbors of $v$ and $C(v)$ is the clustering
coefficient of the vertex $v$.

\subsection{Modularity of a network}
Newman and Girvan~\cite{ng2002} proposed a metric called {\it modularity} that can judge the goodness of a
community detection method. It is based on the concept that random networks do not form strong communities.
Given a  partition of a network into $M$ groups, let  $C_{i j}$ represent the fraction of total links starting
at a node in group $i$ and ending at a node in group $j$. Let $a_i = \sum_{j} C_{i j}$ corresponds to the
fraction of links connected to subgroup $i$. Under random connections, the probability of links that begin at a
node in $i$ is $a_i$, and the probability of links that end at a node in $j$ is $a_j$. Thus, the expected number
of within-community links of group $i$ (i.e., links between nodes in group $i$) is $a^2_i$. The actual fraction
of links within each group $i$ is $C_{ii}$. Therefore, a comparison of the actual and expected values, summed over all
the partitions gives us the modularity, which is the deviation of the partitions from the perfectly random case: $Q = \sum(C_{ii} -a^2_i)$.
Generally, the higher the modularity, the better is the estimation of the
correct community structure in the network.

\subsection{Normalized mutual information (NMI)}
The problem of comparing different community detection approaches can be reduced to comparing how good the
partitions produced by each of the approaches are when compared against the ground-truth. One way to test this
goodness would be to compute the Normalized Mutual Information (NMI)~\cite{nmi, nmi_1}. Let $C$ be the confusion matrix. Also let 
$N_{ij}$ (elements of the confusion matrix $C$) be the number of nodes in the
intersection of the original community $i$ and the generated community $j$. If $C_A$ denotes the number of the
communities in the ground truth, $C_B$ the number of the generated communities by an approach, $N_i$ the sum of
row $i$, $N_j$ the sum of column $j$, and $N$ the sum of all elements in $C$, then the NMI score between the
ground truth partition $A$, and the generated partition $B$ can be computed as shown in the following equation.
\begin{equation}
 NMI(A,B)=\frac{-2\sum\limits_{i=1}^{C_A}\sum\limits_{j=1}^{C_B}N_{ij}log\frac{N_{ij}N}{N_iN_j}}
{\sum\limits_{i=1}^{C_A}{N_i}log\frac{N_i}{N}+\sum\limits_{j=1}^{C_B}{N_j}log\frac{N_j}{N} }
\end{equation}
The values of NMI range between 0 and 1 where 0 refers to no match with the ground truth and 1 refers to a perfect match. 

All the notations
that are used in the paper are
tabulated in Table~\ref{notation}.

\begin{table}[t!]
 \centering
\caption{Notations used and their descriptions}\label{notation}
\small{
\scalebox{0.7}{
\begin{tabular}{|c|c|c|}
\hline
Notation & Name & Description/ Functionality\\\hline
$\mu$ & Mixing parameter & The ratio of the external connections of a node to its total degree\\
      &  of the LFR graph& \\\hline
$\phi$ & Sensitivity & The number of constant communities to the total number of vertices\\
       &of a network & \\\hline
$\xi$ & Relative size & Number of vertices in the constant community to the total number \\
       &of the constant community& of vertices in the network\\\hline
$D(v)$ & Degree & Degree of vertex $v$\\\hline
$IN(v)$ &Internal neighbor & Neighbors of vertex $v$ internal to the community of $v$\\\hline
$EN(v)$ & External neighbor & Neighbors of vertex $v$ outside the community of $v$\\\hline
$ENG(v)$& External neighbor & A set of elements each of which represents the number of external neighbors \\
    &group of $v$ & of $v$ distributed among the communities other than that of $v$   \\\hline
$DENG(G)$ &Distinct external neighbor & The union set of the $ENG$ of all vertices in the network\\\
       & group of the graph $G$& \\\hline
$Rank_i(ENG(v))$& Rank of the $i$th entry & Rank of the $i$th entry of the $ENG(v)$ obtained by sorting the elements\\
            & of the $ENG(v)$ & of the set $DENG(G)$ in ascending order\\\hline
$\Theta$ & Strength & Ratio of the edges internal to the constant community to the edges\\\
         & of a constant community & external to the constant community\\\hline
$\theta(v)$ & Internal strength & Ratio of the number of internal neighbors to the number of external neighbors of $v$ \\
             & of vertex $v$ & \\\hline
$\Omega(v)$& Relative permanence  & It indicates the propensity of the vertex $v$ to stay in a single community despite\\
         & of vertex $v$ & any vertex perturbations or different algorithms used \\\hline
$m_q$ & Mean modularity & Average of the modularity values obtained from the different permutations \\ & & of the input sequences\\\hline
$\sigma_q$ &Variance of the  & Variance of the modularity values obtained from the different permutations \\
       & modularity & of the input sequences\\\hline
$C(v)$ & Clustering coefficient of a vetrex& Clustering coefficient of a vertex\\\hline
$\tilde{C}$ & Avg. clustering coefficient& Average clustering coefficient of the network\\
      & of a network & obtained by averaging the clustering coefficient of all vertices\\\hline
$k$& Degree of a vertex & Degree of a vertex \\\hline
 $P(k)$& Cumulative degree distribution & Fraction of
vertices having degree greater than or equal to $k$ \\\hline

 $H(f_i,f_j)$& Hamming distance between  binary&  The hamming distance between two binary vectors $f_i$ and $f_j$ of equal length\\
        & vectors $f_i$ and $f_j$ & is the number of positions at which the corresponding symbols are different.\\\hline

\end{tabular}}
}
\end{table}

\section*{Comparing properties of the real-world networks}  The results in this section demonstrate that the real-world networks in our test
suite possess characteristics such as power-law degree distribution and high average clustering coefficient.
However, we also see that when comparing with the data in Figure 2 and Figure 3 in the main document, the above characteristics do
not necessarily guarantee that the network has strong community structure.

\subsection{Degree distribution}
An important characteristic of many real-world networks is that they exhibit power-law degree distribution~\cite{bara}. That is,
if the fraction of nodes having degree greater than or equal to $k$ is $P(k)$, then $P(k) \approx ck^{-\gamma}$, where $c$ is a constant and
the value of
$\gamma$ is generally between $2 \le \gamma \le 3$.   Figure S1 shows that all the networks in our test-suite exhibit power law
distribution; however not all of them are found to possess constant communities (see Figures 2 and 3 in the main document). Moreover, the
slope
of the curve does not provide an indication of the presence of constant communities. For example, Email and Polbook have nearly similar
slopes, but Email does not exhibit any constant communities, while Polbook has about three large constant communities.

\begin{figure}[!h]
 \centering
\includegraphics[scale=0.3]{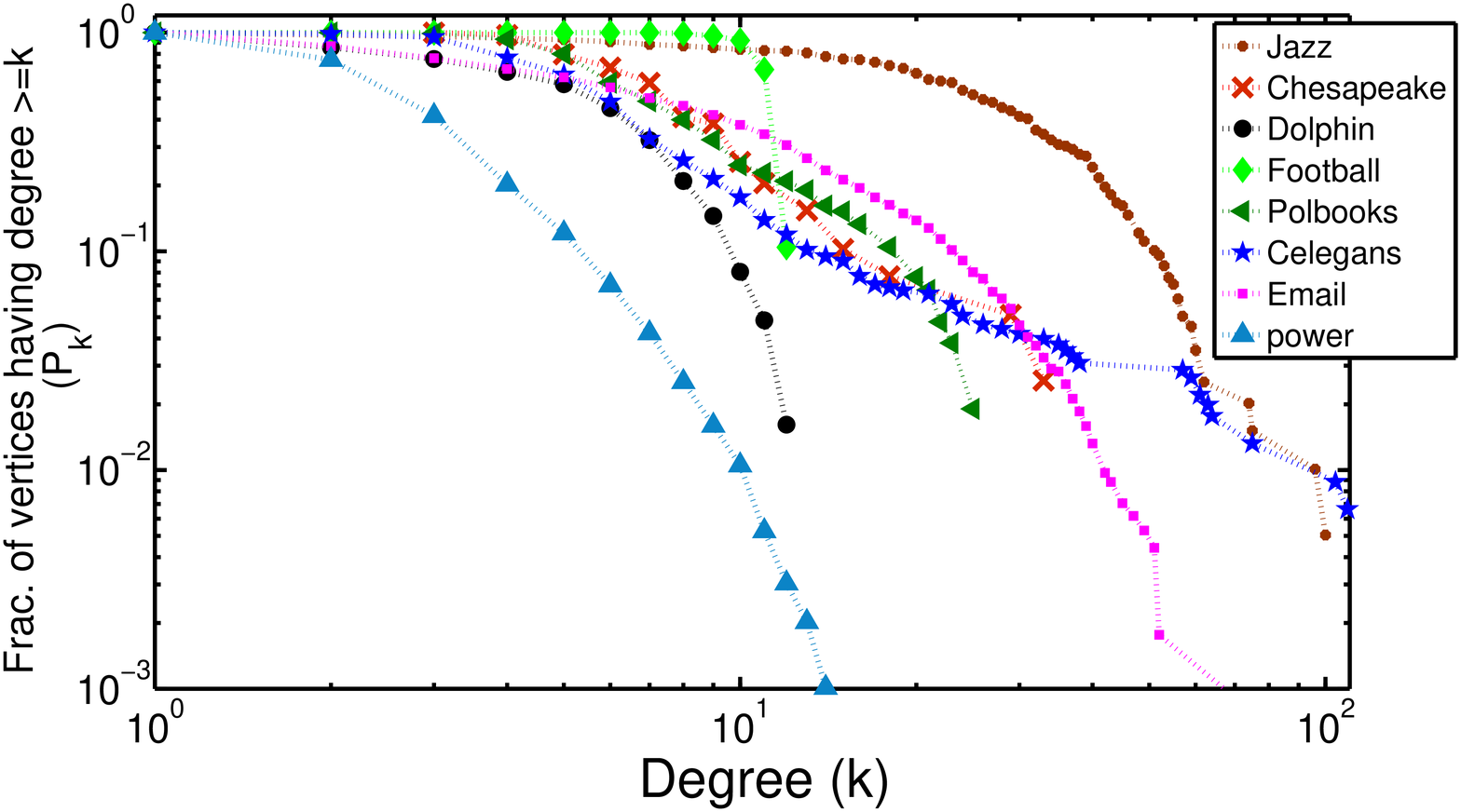}
\caption{ [S1] {\bf Cumulative degree distributions of the real-world networks.} The networks, regardless of the number of constant
communities
present, exhibit power-law degree distribution.}\label{deg}
\end{figure}

\subsection{Average clustering coefficient}
 We computed the average clustering coefficient for a network with $n$ vertices as $\tilde{C}=(1/n)\times \sum_{i=1}^{i=n}C(i)$. We created
the random graphs using the Erdos-Reyni
graph~\cite{er} generator in MatlabBGL with the probability of connection between the nodes set chosen such that the
number of edges is close to the original networks. Table~\ref{er} compares the clustering coefficients obtained from
the original graph and the corresponding Erdos-Reyni (ER) graph. The values indicate that the networks in the set are
indeed more densely packed than the random graphs.

\begin{table}[!h]
 \centering
\small{
\caption{{\bf Average clustering coefficients of the real-world networks}. The values are higher than those
obtained from a random network of nearly the same size.}\label{er}
\begin{tabular}{|c|c|c|c|}
\hline

{Name} & { $\tilde{C}$}  & {Vertex ; Edges } & {$\tilde{C}$ } \\
 & { (Original)} & { of ER }  & {(ER)}

 \\\hline\hline
Jazz & 0.6174 &198; 2042 & 0.0407\\
Chesapeake & 0.4502 & 39; 340 & 0.2134 \\
Dolphin & 0.2589  &  62 ; 156 & 0.0183 \\
Football & 0.4032 &115; 1226 & 0.1026\\
Polbooks & 0.4875 & 105; 426 & 0.0443\\
Celegans & 0.6464 & 453; 2048 & 0.0102\\
Email & 0.2201 & 1133; 5170 & 0.0021\\
Power & 0.0801 &4941; 6386 & 4e-04\\\hline
\end{tabular}
} 
\end{table}

\begin{table}[!h]
 \centering
\caption{Comparison between the constant communities obtained from Louvain and CNM algorithms using NMI}\label{nmi_real}
\begin{tabular}{|c|c|c|c|c|c|c|c|c|}
\hline
\textbf{Networks} & Jazz & Chesapeake & Dolphin & Football & Polbooks & Celegans & Email & Power \\\hline
\textbf{NMI} &  0.8856& 0.8429 & 0.8663 & 0.8765 & 0.8950 & 0.9232& 0.8103& 0.8097\\\hline
\end{tabular}
\end{table}

\section*{Comparing constant communities obtained from two algorithms}
The primary intuition behind constant community is that these sub-modules are invariant under any circumstance,
i.e., across any ordering of the vertices or any non-deterministic, optimized algorithm used to detect the
community structure from the network. We have judged the invariability of the structure of the constant
community for two algorithms -- Louvain and CNM. The comparison of the constant community structure for these
two methods using NMI is tabulated in Table~\ref{nmi_real}. For all the cases, the NMI value is greater than
0.80 which proves to be reasonably standard~\cite{nmi_1, nmi_2} indicating the high overlap between the partition
structures of the detected constant communities from two different algorithms. This follows our initial claim that
the constant communities are nearly invariant across different community detection algorithms.

\section*{Feature overlaps of constant communities}

We conduct constant community analysis of PhoNet and compute the average hamming distance between the feature vectors of the constituent
members of the community. We report in Figure S2 the average hamming distance ($H(f_i,f_j)$, see Table~\ref{notation}) versus the size of
the communities and compare the results
with randomly constructed same-sized groups of phonemes showing that the constant communities of PhoNet are far from being arbitrary. In
addition, we observe that collapsing the constant communities produce communities that are functionally dilute and at times could be quite
relevant for certain applications. Note that the larger the size, the lesser the feature overlap since a large group would have higher
chances to admit more feature variations.  

\begin{figure}
\centering
\includegraphics[scale=0.3]{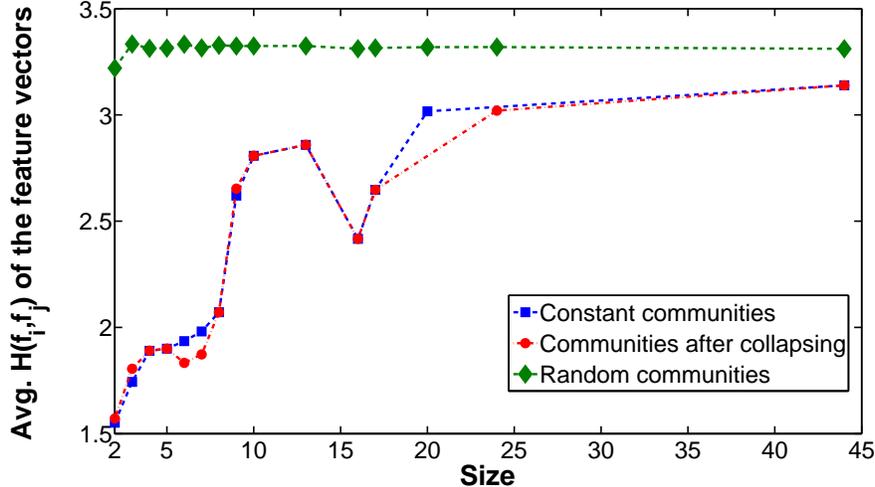}
\caption{ {\bf [S2] Feature overlap of constant communities, communities after collapsing and random communities
of different size}. X-axis denotes the size of the community and y-axis denotes the average pair-wise
hamming distance of the feature vectors.}\label{phonet}
\end{figure}

\begin{figure}
\centering
\includegraphics[scale=0.45]{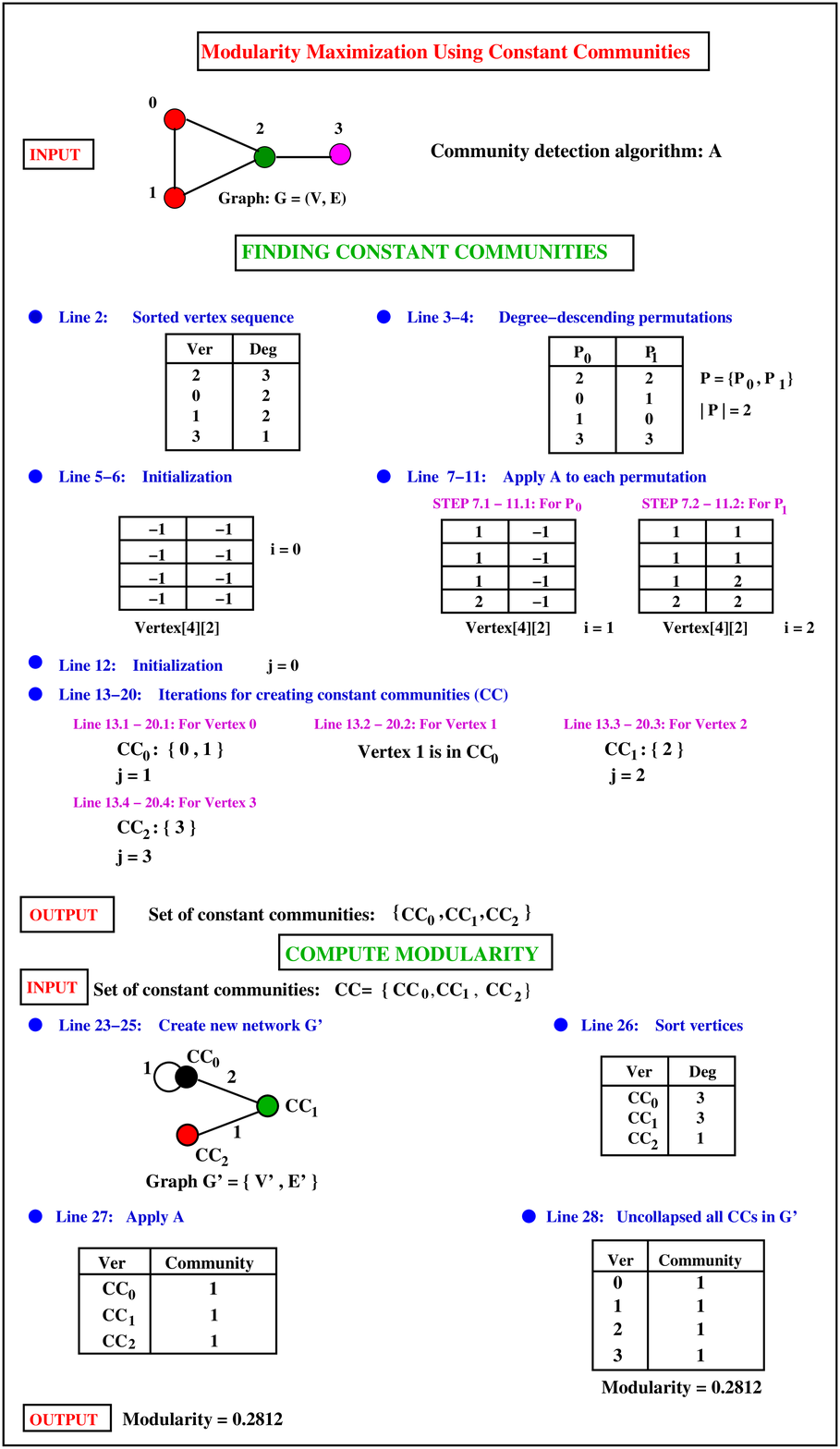}
\caption{\bf{ [S3] Schematic diagram of the proposed algorithm (Algorithm 1) for modularity maximization using constant
communities.}}\label{algo_pic}
\end{figure}

\section*{Modularity Maximization Using Constant Communities}
We provide a schematic diagram of the Algorithm 1 in Figure S3. The process consists of two steps: first, the
pre-processing step for finding constant communities, and then enhancing the performance of the community
detection algorithm using the detected constant communities. Initially, the vertices are ordered according to
their degrees (Line 2 in Algorithm 1). The permutations of the vertex preserve this order, that is, if vertex
$v_i$ is placed before $v_j$ in the sequence then $degree(v_i) \ge degree(v_j)$ (Lines 3 and 4). We then compute
the communities obtained for each permutation $i$ (Lines 7-11). The constant communities constitute those vertices which are
always assigned together (Lines 13-20).

The second step consists of collapsing the constant communities into a single super-vertex (Lines 23-25). The
edges from the super-vertex are weighted to reflect the number of connections from that vertex to the rest of
the network. Self loops are also included to represent internal connections within the constant communities. The
network with the super-vertex is called the collapsed network. We again permute the vertices according to their
descending degree (Line 26) and find the communities (Line 27). We then unfold the super-vertices back to their
constituent vertices (Line 28), and compute the modularity on the network.

We compute the variance in the modularity values and the arithmetic mean, and compare the results of the computation with and without using
constant communities in the pre-processing step. The results of Table I in the main
document show that pre-processing leads to higher modularity values on average as well as less variance among the
results.\\\\
{\bf Algorithm 1:} Modularity Maximization Using Constant Communities\\
  {\bf Input:}  A network (graph) $G=(V,E)$; Community Detection Algorithm $A$.\\
  {\bf Output:} Set of Constant Communities ${CC_1}$, \ldots${CC_k}$; Modularity $Q$
\begin{algorithmic}[1]
  
  \Procedure{Finding Constant Communities}{}
  \State{Sort vertices in $V$ in degree descending order}
  \State {Apply degree preserving permutation $P$ to vertices such that degree($v_i$) $\ge$ degree($v_{i+1}$) \hspace*{.2 in} in $P$.}
  \State {$|P|$ is number of degree preserving permutations applied.}
  \State{ Initialize  array $Vertex[|V|][|P|]$ to -1}   \Comment{$Vertex[|V|][|P|]$ will store the community \hspace*{.2 in} membership of
the vertices in
	  each permutation}
 
  \State{Set $i=0$}  \Comment{This variable indicates the permutation index}
  
   \ForAll {$P_i \in P$} \Comment{Detect community memberships of the vertices in each \hspace*{.2 in} permutation using $A$ and store them
in
$Vertex$}
    \State {Apply algorithm $A$ to find the communities of the permuted network $G_{P_{i}}$}
    \If {Vertex $v$ is in community $c$} 
       \State {$Vertex[v][i]=c$} \Comment{Vertex $v$ in permutation $P_i$ belongs to community c after \hspace*{.7 in} applying $A$ to
$P_i$}
    \EndIf 
      \State {$i=i+1$}
    \EndFor
    
    \State{Set $j=0$}  \Comment{This variable indicates the index of the constant community}
    \ForAll{ $v \in V$} \Comment{Detecting constant communities using the community \hspace*{.2 in} information stored in $Vertex$}
	\If {vertex $v$ is not in a constant community}
	  \State {Create constant community $CC_j$}
	  \State {Insert $v$ to $CC_j$}                 \Comment{All $CC_{j}s'$ are the constant communities}
	    
        \ForAll{ $u \in V \setminus CC_j$}
        
        \If{ $Vertex[v][i] = Vertex[u][i]$, $\forall$ $i=1$ to $|P|$} \Comment {Check for the exact \hspace*{.9 in} matching of community
memberships of
{\it u} and {\it v}}
        \State {Insert $u$ to $CC_j$ }
        \EndIf
        \EndFor
    
      \EndIf
    \State {$j=j+1$}
    \EndFor

\EndProcedure

  \Procedure{Computing Modularity}{} 
  
\State{ Set of constant communities in $CC$}
\ForAll { $CC_j \in CC$}   \Comment{Create intermediate small, weighted network}
  \State {Combine vertices in $CC_j$ into a super-vertex $X_j$}
  \State {Replace edges from $X_j$ to another vertex $X_i$ by their aggregate weight} \Comment{For the \hspace*{.45 in} self-loop, i=j} 
\EndFor
\State {Sort vertices of collapsed network, $G'$, in degree descending order}
\State {Apply community detection method $A$}
\State {Unfold all $X_j$ in $G'$ and compute the modualrity $Q$}
\EndProcedure

\end{algorithmic}

\bibliographystyle{model1-num-names}

\end{document}